\documentclass[journal]{IEEEtran}
\ifCLASSINFOpdf
\else
\fi
\usepackage[cmex10]{amsmath}
\usepackage{ amssymb }
\usepackage[pdftex]{graphicx}
\usepackage{float}
\usepackage{multicol}
\usepackage{lipsum}
\usepackage{ifpdf}
\usepackage{multirow}
\usepackage{epstopdf}
\usepackage{color}
\usepackage{cite}
\usepackage{graphicx}
\usepackage{soul}
\usepackage{balance}
\usepackage{threeparttable}
\usepackage{wasysym}
\usepackage{url}
\usepackage[normalsize]{subfigure}
\usepackage{enumitem}
\usepackage{enumitem}
\usepackage{dblfloatfix}    
\usepackage{afterpage}
\usepackage{changepage} 
\usepackage[hidelinks]{hyperref}
\usepackage{accents}
\usepackage{algorithmicx}
\usepackage{algpseudocode}

\newcommand{\ubar}[1]{\underaccent{\bar}{#1}}

\usepackage{fancyhdr}

{}
\begin{document}
\IEEEpeerreviewmaketitle

\title{Optimal Planning and Operation of Multi-Frequency HVac Transmission Systems}

\author{Quan~Nguyen,~\IEEEmembership{Student Member,~IEEE,}
	and~Surya~Santoso,~\IEEEmembership{Fellow,~IEEE}
	\vspace{-0.5cm}
}


\maketitle

\thispagestyle{fancy}

\begin{abstract}
	Low-frequency high-voltage alternating-current (LF-HVac) transmission scheme has been recently proposed as an alternative solution to conventional 50/60-Hz HVac and high-voltage direct-current (HVdc) schemes for bulk power transfer. This paper proposes an optimal planning and operation for loss minimization in a multi-frequency HVac transmission system. In such a system, conventional HVac and LF-HVac grids are interconnected using back-to-back (BTB) converters. The dependence of system MW losses on converter dispatch as well as the operating voltage and frequency in the LF-HVac is discussed and compared with that of HVdc transmission. Based on the results of the loss analysis, multi-objective optimization formulations for both planning and operation stages are proposed. The planning phase decides a suitable voltage level for the LF-HVac grid, while the operation phase determines the optimal operating frequency and power dispatch of BTB converters, generators, and shunt capacitors. A solution approach that effectively handles the variations of transmission line parameters with the rated voltage and operating frequency in the LF-HVac grid is proposed. The proposed solutions of the planning and operation stages are evaluated using a multi-frequency HVac system. The results show a significant loss reduction and improved voltage regulation during a 24-hour simulation.
\end{abstract}


\IEEEpeerreviewmaketitle

	\vspace{-0.3cm}
\section{Nomenclature}
	\vspace{-0.1cm}
	Subscript $*$ denotes the transmission scheme of the power system:
	\vspace{0.1cm}
	
	\begin{tabular}{p{1.8cm} p{5.8cm}}
		\textit{$s$}          & 50/60-Hz HVac transmission system, \\
		\textit{$l$}          & LF-HVac transmission system, \\
	\end{tabular}\\
	
	Sets:
	\vspace{0.1cm}
	
	\begin{tabular}{p{1.8cm} p{5.8cm}}
		\textit{$\mathcal{N}_*$}          		& set of buses in a grid, \\ 
		\textit{$\mathcal{D}_{*}$}      		& set of transmission lines,    \\	
		\textit{$\mathcal{G}_*$}          		& set of buses connected to generators, \\	
		\textit{$\mathcal{L_*}$}          		& set of non-voltage-controlled buses, \\ 
		\textit{$\mathcal{C_*}$}      	  		& set of buses connected to converters,\\ 	
		\textit{$\mathcal{V}_*$}          		& set of buses connected to converters operating in voltage-controlled mode, \\	
		\textit{$\mathcal{N}_{*}^{sh}$}    		& set of buses with shunt capacitors, 	\\
		\textit{$\mathcal{Q}_{*, k}^{sh}$} 		& set of discrete dispatch of the shunt capacitor at bus $k$, 	
	\end{tabular}\\
	
	Parameters:
	\vspace{0.1cm}
	
	\begin{tabular}{p{1.8cm} p{5.8cm}}
		\textit{$P_{s}^{load}, Q_{s}^{load}$}   & loads at buses in HVac grid $s$,\\	
		\textit{$g_{*}, b_{*}$}   & line series conductance and susceptance,\\	
		\textit{$V_{*}^{sch}$}   				& scheduled voltage magnitude at voltage-controlled buses,\\	
		\textit{${\ubar{V}}, {\bar{V}}$}		& lower and upper load voltage limits,\\	
		\textit{${\bar{I}_{*}}$}				& maximum line current,\\		
		\textit{${\bar{P}_{*}}$}				& maximum line real power,\\	
		\textit{$\boldsymbol{Y}_{*}$, $\boldsymbol{G}_{*}$, $\boldsymbol{B}_{*}$}			& admittance, conductance, and susceptance matrices,\\	
		\textit{$\alpha_1$, $\alpha_2$, $\alpha_3$} 		& weighting coefficients,\\
	\end{tabular}
	
	Variables:
	\vspace{0.1cm}
	
	\begin{tabular}{p{1.8cm} p{5.8cm}}
		\textit{$V_l, F_l$}    					& optimal operating voltage and frequency of LF-HVac transmission system $l$,\\
		\textit{$e_*, f_*$}    					& real and imaginary parts of voltages,\\
		\textit{$P_*, Q_*$}         			& injected power into grid from buses,	\\\
		\textit{$P_*^{gen}, Q_*^{gen}$}         & generator dispatch,\\
		\textit{$P_*^{conv}, Q_*^{conv}$}       & power from/to BTB converters,	\\	
		\textit{$Q_*^{sh}$}        			    & injected reactive power into grid from shunt capacitors at 1 pu.
	\end{tabular}

\section{Introduction}
	\vspace{-0.05cm}
	In spite of the increasing penetration of distributed energy resources in the distribution system, the transmission system plays an ever-important role in transporting bulk power over long distances.  Proposed supergrids in Europe and Asia, as well as, the increasing number of offshore wind farms provide a motivation to seek an improved and alternative bulk power transmission scheme \cite{Gordon_1,Barthold_1,Movellan_1}. Compared to the conventional 50/60-Hz high-voltage alternating current (HVac) transmission systems, converter-based high-voltage direct current (HVdc) systems have shown unique benefits of unrestricted point-to-point bulk-power transfer capability over long distances, reduced line losses, and narrower right of way. However, reliable HVdc operation is confronted by the immature dc circuit breakers in fault clearing. This operational challenge considerably impedes the feasibility of replicating multi-point interconnection capability of HVac systems.
	
	Considering the limitations of conventional HVac and HVdc technology, growing attention has turned to an alternative solution for bulk-power transmission, which is called low-frequency HVac (LF-HVac). LF-HVac offers the advantages of the two existing technologies, such as high power-carrying capability over long distance, straightforward ac protection system, and the flexibility of multi-terminal networks \cite{Funaki_1,Chen_1,PSERC_1}. A significant reduction in reactance at low frequency also improves voltage profiles and system stability \cite{Tuan_1, Rosewater_1, Johnny_1}.  	Similar to HVdc systems, an LF-HVac system requires power converters for connection to a conventional 50/60-Hz HVac system, which forms a multi-frequency power system. Converter topologies proposed to perform the connection include ac/ac cycloconverters, ac/ac modular multi-level converters (MMC), and ac/dc/ac back-to-back (BTB) converters. In this paper, the BTB configuration is chosen as the converter topology because it allows a smaller filter size while offering full power and frequency control capabilities \cite{Ruddy_2, Tang_1,Quan_0}. Existing control techniques and equipment in HVdc systems can also be directly adopted for BTB converters.

	\begin{figure*}[b!]
		\centering
		\includegraphics[width = 1.6\columnwidth] {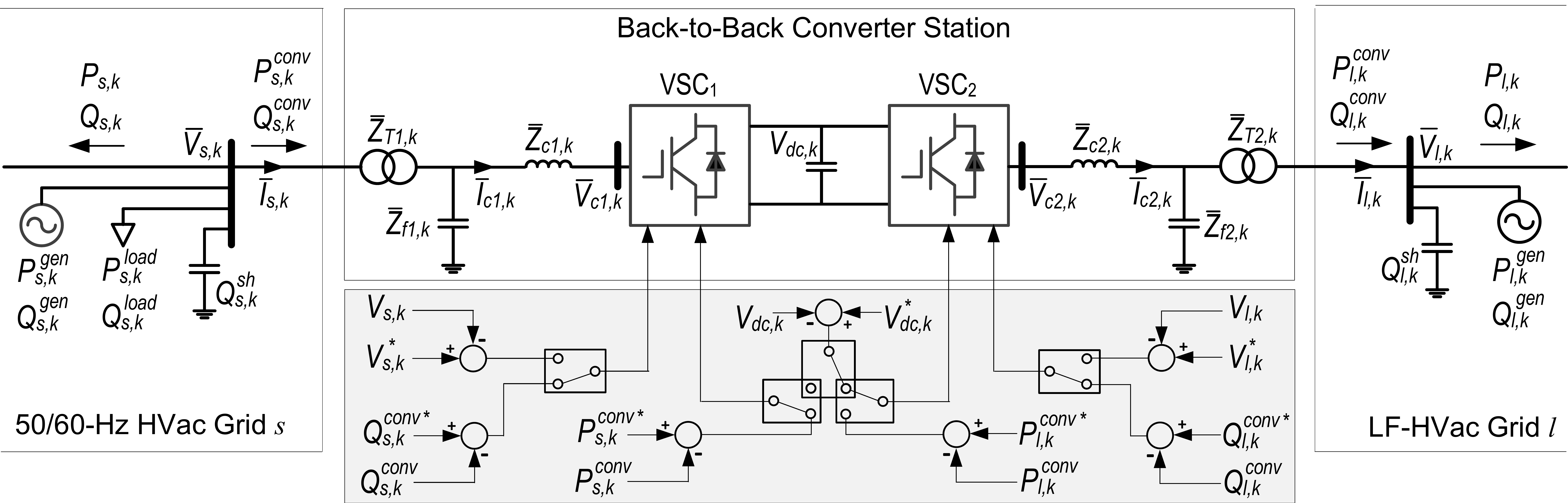}	
		\vspace{-0.2cm}
		\caption{A BTB converter is used to connect an LF-HVac grid to a 50/60-Hz HVac grid: the system configuration, the interface between the two grids, and the main control blocks. }
		\vspace{-0.2cm}	
		\label{fig:Back-to-Back_Converter_Control}
	\end{figure*}

	Several research has been done to evaluate the potential applications and performance of an LF-HVac grid as a standalone system as well as a part of a generalized multi-frequency system. Practical 16.7 Hz LF-HVac transmission for European offshore wind farms are proposed in \cite{Tom_1,Erlich_1,Ruddy_1}. In \cite{Ruddy_2,Ruddy_3}, a laboratory-scale point-to-point LF-HVac line is designed and simulated using a real-time hardware-in-the-loop platform to verify the concept and control capability for offshore applications. Since an LF-HVac grid is designed to transfer bulk power, minimizing its losses leads to a significant total loss reduction in the entire system. In \cite{Ruddy_4}, a comparison of power losses and costs between HVdc and LF-HVac transmission lines with the length up to 250 km and at a frequency range from 10 to 16.7 Hz are shown. A selection of an optimum frequency based on transmission range is also included. However, the voltage of the LF-HVac grid, real power transfer, and reactive power support from BTB converters, which greatly affect system losses and the optimal frequency, are not discussed. Therefore, a system-wide model is mandatory to accurately evaluate suitable voltage levels and operating frequencies for a multi-terminal LF-HVac grid in a multi-frequency HVac system. In \cite{Quan_3}, power flow formulation and solution for such a multi-frequency system are proposed. However, the chosen rated voltage, operating frequency, and scheduled power transfer do not guarantee an optimal performance in terms of losses and voltage regulation. 
	
	As an extension of the existing works on the emerging LF-HVac transmission technology, the main contributions of this paper are:
	\begin{itemize}[leftmargin=*]
		\item A loss study and comparison between LF-HVac and HVdc transmission schemes, with respect to scheduled power transfer, system voltage, and operating frequency. The results of this comparison show the need of a generalized system-wide OPF model to achieve an optimal operation in a multi-frequency system.
		\item Multi-period multi-objective OPF formulations for planning and operation stages of a multi-frequency HVac system. During the planning stage, a suitable voltage for the LF-HVac grid is determined to achieve minimum losses. During the operation phase, the actual optimal frequency and dispatch from generators, shunt capacitors, and converters are determined given real-time load data, subjected to comprehensive operational constraints of all ac grids and converters.
		\item An scalable and effective solution approach to handle the variations of transmission line parameters with the rated voltage and operating frequency in the power flow constraints of the LF-HVac grid.
	\end{itemize}

\section{Operation of Back-to-Back Conveters in a Multi-Frequency HVac Power System}\label{sec:BTB_Converter_Modeling}
	This section briefly describes the operation of BTB converters in a multi-frequency HVac system. Fig. \ref{fig:Back-to-Back_Converter_Control} shows the model and control of each BTB converter, which consists of two VSC denoted as $\textup{VSC}_\textup{1}$ and $\textup{VSC}_\textup{2}$. These converters have identical structures and electrical components such as a transformer, a phase reactor, and switching devices. These two converters share a common dc-link capacitor with a constant dc voltage that allows decoupled operation of $\textup{VSC}_\textup{1}$ and $\textup{VSC}_\textup{2}$. 
	
	When $n$ BTB converters are used to connect HVac grid $s$ and LF-HVac grid $l$, the following operating modes are applied to regulate the power transfer, dc-link voltage, and/or ac terminal voltage magnitude of $n$-1 BTB converters \cite{Quan_3}:

	\begin{itemize}[leftmargin=*]
		\item Converter $\textup{VSC}_\textup{1}$, which is connected to HVac grid $s$, is set to regulate real and reactive power (PQ mode) or real power and ac terminal voltage magnitude (PV mode). This means that $P_{s,k}^{conv}$ and $Q_{s,k}^{conv}$ in PQ mode or $P_{s,k}^{conv}$ and $V_{s,k}$ in PV mode are known parameters.
		\item  Converter $\textup{VSC}_\textup{2}$, which is connected to LF-HVac grid $l$, is set to regulate the dc-link voltage $V_{dc,k}$ and to control either reactive power (Q$\textup{V}_\textup{dc}$ mode) or the ac terminal voltage magnitude (V$\textup{V}_\textup{dc}$ mode). This implies that either $Q_{l,k}^{conv}$ in Q$\textup{V}_\textup{dc}$ mode or $V_{l,k}$ in V$\textup{V}_\textup{dc}$ mode is a known quantity while $P_{l,k}^{conv}$ is always unknown and needs to be determined.
	\end{itemize}
	This approach does not apply to a BTB converter that is connected to the slack bus in LF-HVac grid $l$. At the $\textup{VSC}_\textup{2}$ side of this slack converter, $P_{l,sl}^{conv}$ and $Q_{l,sl}^{conv}$ vary to account for the losses in the LF-HVac grid and control voltage magnitude $V_{l,sl}$ at the slack bus. The voltage angle of the slack bus is also considered as the reference angle for other buses in grid $l$. At the $\textup{VSC}_\textup{1}$ side of this BTB converter, only the reactive power $Q_{s,sl}^{conv}$ is controlled. The real power $P_{s,sl}^{conv}$ is unknown since it depends on the unknown real power $P_{l,sl}^{conv}$, Joule losses due to the resistive elements of transformer and reactor impedances $\bar{Z}_{T1}$, $\bar{Z}_{c1}$, $\bar{Z}_{T2}$, and $\bar{Z}_{c2}$, and switching losses in $\textup{VSC}_\textup{1}$ and $\textup{VSC}_\textup{2}$ converters.

	\vspace{-0.2cm}
\section{Dependence of Losses on Voltage, Frequency, and Reactive Power Support in LF-HVAC Grid}	
\begin{figure}[t!]
		\centering
		\subfigure[]{
			\includegraphics[width = 0.9\columnwidth]{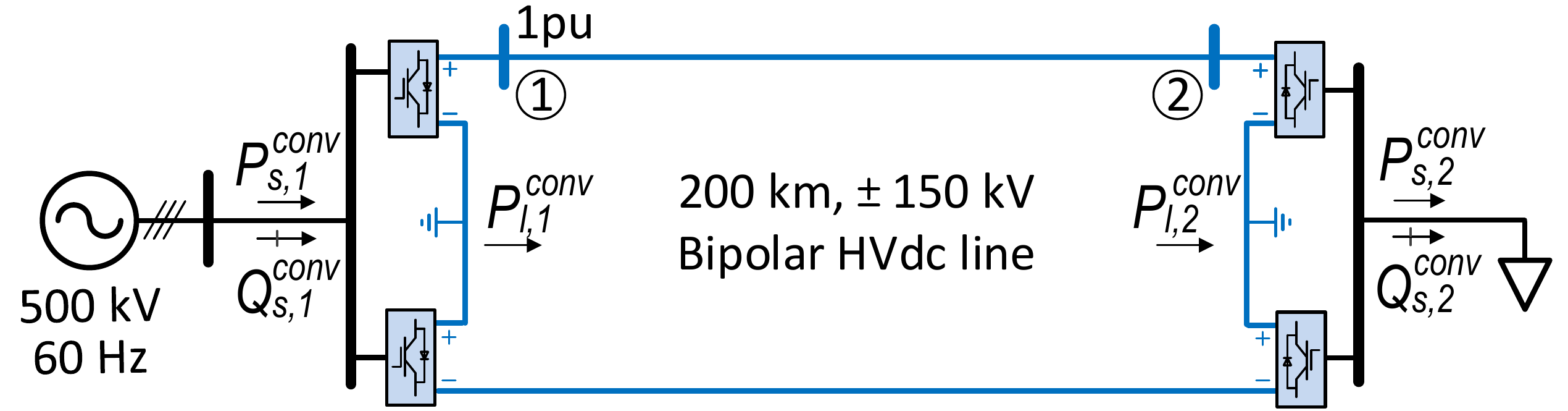}
			\label{fig:Bipolar_HVdc_Loss_Comparison}}
		\quad
		\subfigure[]{
			\includegraphics[width = 1.01\columnwidth]{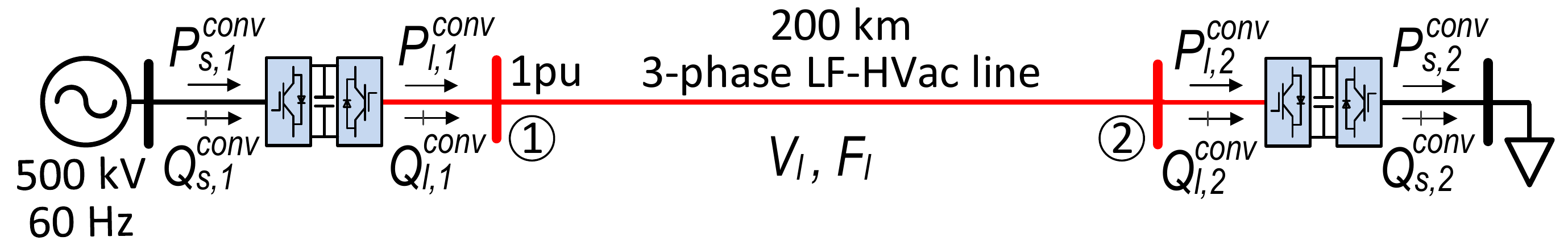}
			\vspace{-0.2cm}
			\label{fig:LF-HVac_Loss_Comparison}}
		\vspace{-0.5cm}
		\caption{a) A 200-km line employing bipolar HVdc (adopted from BorWin1 system in Germany \cite{BorWin1_1}), and b) an alternative configuration employing LF-HVac transmission.}
		\label{fig:HVdc_LF-HVac_Loss_Comparison}
	\end{figure}

	\vspace{-0.05cm}
	This section presents a comparison on losses between two 200-km transmission lines employing HVdc and LF-HVac schemes, as shown in Fig. \ref{fig:HVdc_LF-HVac_Loss_Comparison}, under different scheduled power transfer, system voltage, and operating frequency. 
	
	The HVdc grid shown in Fig. \ref{fig:Bipolar_HVdc_Loss_Comparison} is operated at $\pm$150 kV and in bipolar topology. The number of VSCs is thus equal to that in the LF-HVac grid. The power transfer and losses of a bipolar configuration is similar to a double circuit of monopolar 150-kV HVdc but with a significantly less ground current. The LF-HVac grid shown in Fig. \ref{fig:LF-HVac_Loss_Comparison} is operated at rated voltage $V_l$ and frequency $F_l$. The voltage at the sending-end buses, i.e. Bus 1, in both HVdc and LF-HVac grids is kept at 1 pu. In the base case, the two lines deliver same desired amounts of real power $P_{s,2}^{conv}$ from an ideal voltage source in a 60-Hz HVac grid to a remote load. Several assumptions and insights of the comparison are as follows:
	\begin{itemize}[leftmargin=*]
		\item The electric components such as transformers, phase reactors, and switching devices in all VSC stations are identical.
		\item Reactive power $Q_{s,1}^{conv}$ withdrawn from the ideal 60-Hz voltage source and $Q_{s,2}^{conv}$ consumed by the load, which have similar effects in system losses, are set to zero for simplicity.
		\item VSC converters at Bus 1 and Bus 2 in the LF-HVac grid can supply and absorb reactive power from the grid.
	\end{itemize}
	The comparison is conducted using the PSCAD/EMTDC time-domain simulation program and verified by a power flow tool for multi-frequency HVac - HVdc systems reported in \cite{Quan_3}. 
	
\begin{figure}[t!]
		\centering
		\includegraphics[width = 0.98\columnwidth] {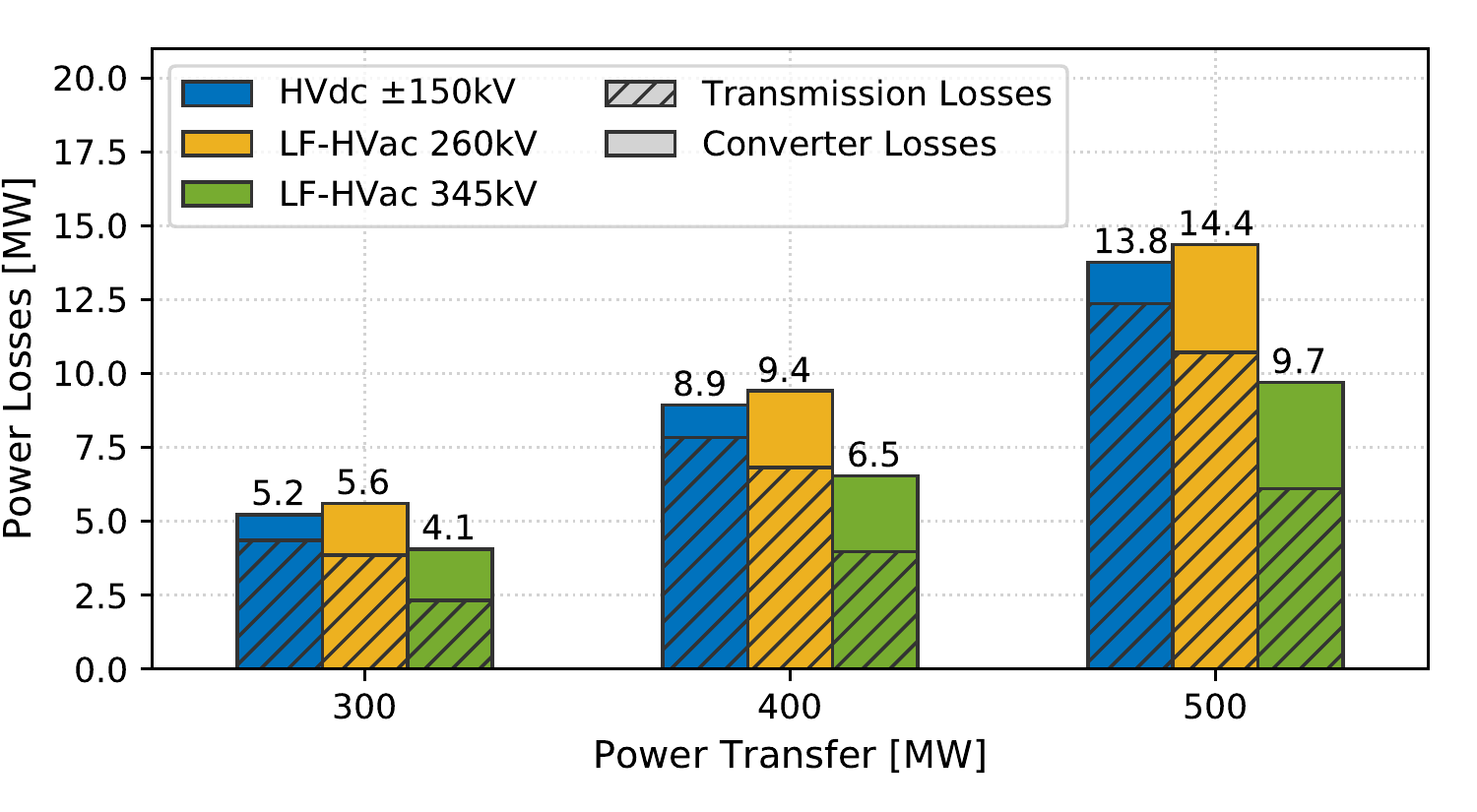}	
		\vspace{-0.5cm}
		\caption{A comparison of system losses varying with real power transfer and operating voltage.}	
		\vspace{-0.1cm}		
		\label{fig:Loss_Ptransfer}
	\end{figure}

	Fig. \ref{fig:Loss_Ptransfer} shows the total system losses, including transmission loss and converter losses in the HVdc and LF-HVac grids at different power transfer and rated voltages. The operating frequency of the LF-HVac line is fixed at 10 Hz. It is expected that a higher system voltage and/or a lower power transfer results in a lower transmission loss. More importantly, it is shown that the transmission losses of the LF-HVac transmission lines are comparable or less than that of the HVdc transmission line. In this study, an LF-HVac voltage of 260 kV is chosen because the corresponding line-to-neutral voltage is equal to the 150 kV line-to-ground HVdc voltage. Another LF-HVac voltage of 345 kV is chosen because it is close to the equivalent 300 kV line-to-line voltage of the bipolar $\pm$150-kV HVdc system.

	\begin{figure}[b!]
		\centering
		\includegraphics[width = 0.97\columnwidth] {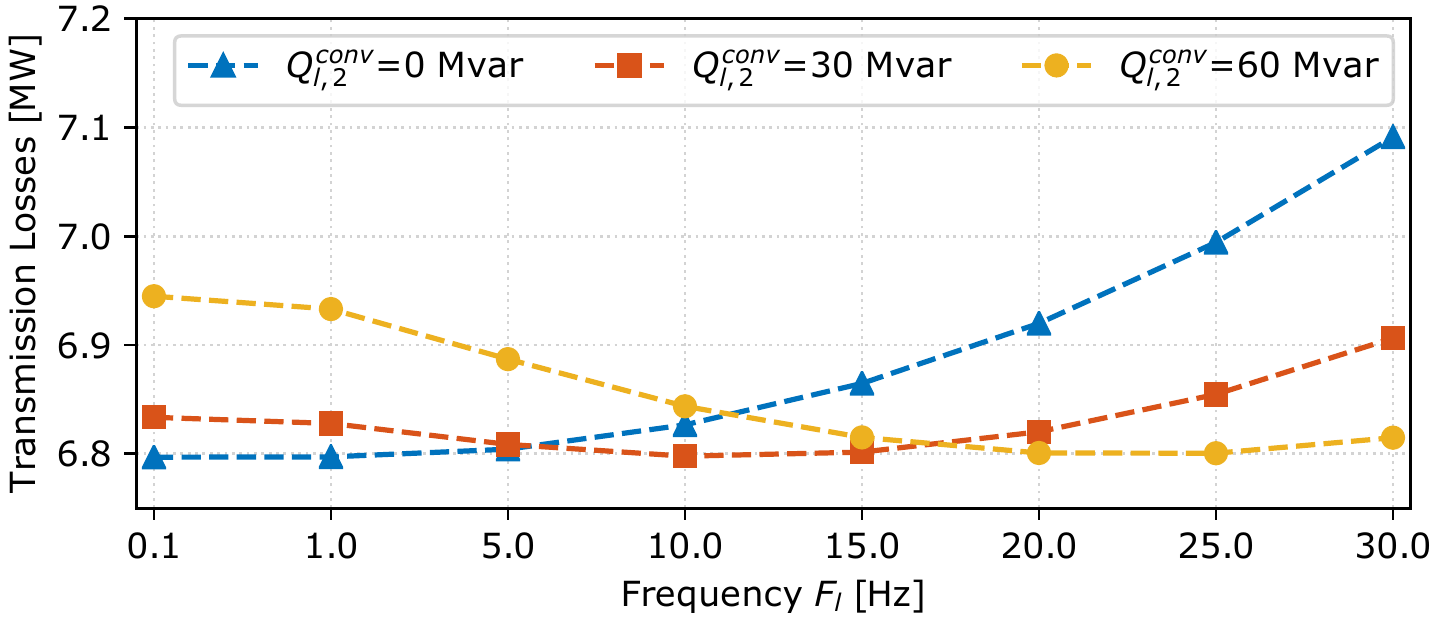}	
		\vspace{-0.3cm}
		\caption{A comparison of transmission losses varying with the operating frequency.} 
		\label{fig:Loss_Frequency}
	\end{figure}
	\begin{figure}[t!]
		\centering
		\vspace{-0.2cm}		
		\includegraphics[width = 0.97\columnwidth] {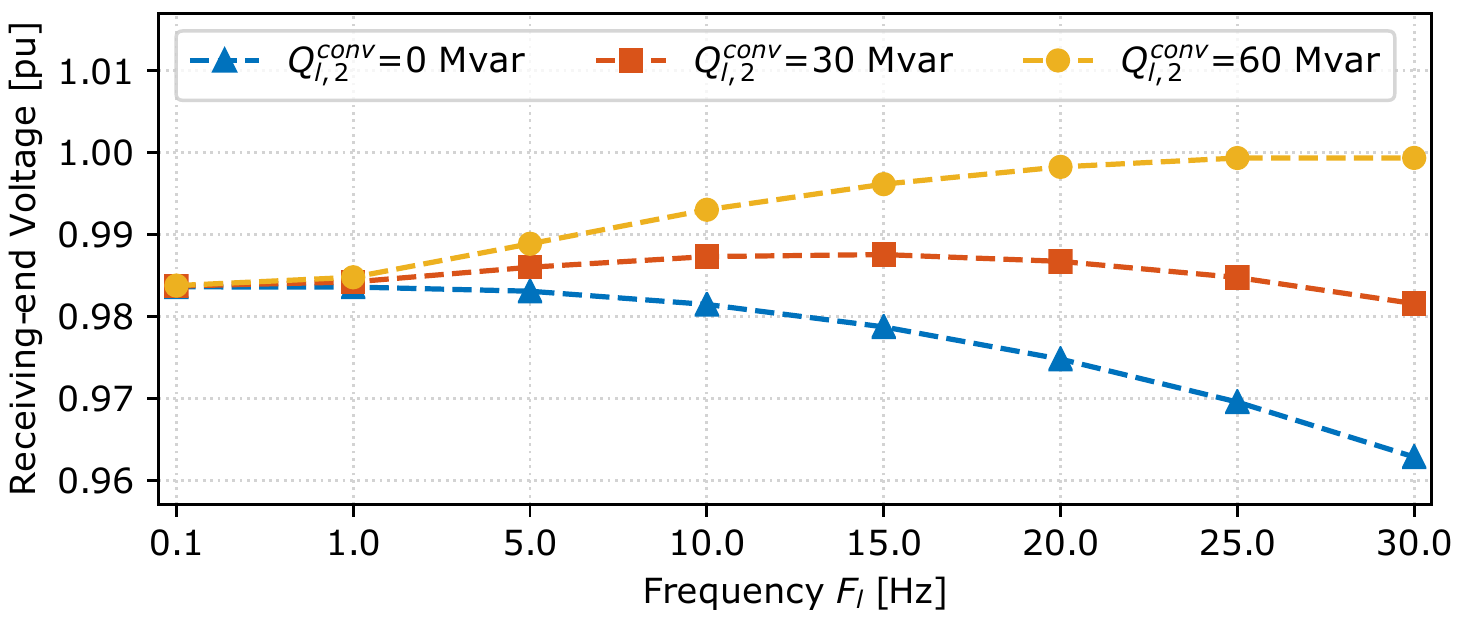}	
		\vspace{-0.3cm}
		\caption{A comparison of receiving-end voltage of the LF-HVac grid varying with the operating frequency. The receiving-end voltage of the HVdc grid shown in Fig. \ref{fig:Bipolar_HVdc_Loss_Comparison} at a similar power transfer $P_{s,2}^{conv}$ = 400 MW schedule is 0.981 pu.}	
		\label{fig:Voltage_Frequency}
	\end{figure}

	Fig. \ref{fig:Loss_Frequency} shows the dependence of the transmission loss and receiving-end voltage of the LF-HVac line on the operating frequency when $V_l$ = 260 kV and $P_{s,2}^{conv}$ = 400 MW. It is shown that for different reactive power schedules at the receiving end, the minimum transmission loss occurs at different frequencies. In this case, when the reactive power injected to the LF-HVac line from Bus 2 is 0, 30 , and 60 Mvar, the optimal operating frequency that results in the lowest loss is 0.1 Hz (the minimum analyzed frequency), 10 Hz, and 25 Hz. In general, the optimal frequency is the one at which the combination of  the supplied and consumed reactive power from line capacitance, line reactance, and converter results in the smallest line current. It is also important to note that VSC converters might increase the supply or absorption of reactive power from the line to reduce transmission losses. However, such a decision might increase converter losses and thus the total system losses.
	
	Fig. \ref{fig:Voltage_Frequency} shows the receiving-end voltage of the LF-HVac grid corresponding to the analysis shown in Fig. \ref{fig:Loss_Frequency}. It can be seen that the optimal frequency that results in the best voltage regulation without reactive power support from the converter at the receiving end, i.e. $Q_{l,2}^{conv}$ = 0 Mvar, is 0.1 Hz. However, when $Q_{l,2}^{conv}$ is 30 Mvar or 60 Mvar, the optimal frequency in terms of voltage regulation is 15 Hz and 25 Hz, respectively. 
	
	The above results show that rated voltage and operating frequency of the LF-HVac grid as well as the scheduled real and reactive power transfer significantly affect the transmission loss and voltage regulation. Due to the nonlinear characteristic of power flow model, it is not straightforward to determine exactly their best combination, i.e. the optimal operating point, that results in lowest system losses or best voltage regulation. Therefore, to achieve a successful and optimal operation of a multi-frequency HVac system, it is important to determine a suitable voltage for the LF-HVac grid during the planning phase as well as an optimal coordination between control resources in the conventional 50/60-Hz HVac grid, LF-HVac grid, and the BTB converters connecting the grids during the operation period.
	

\vspace{-0.2cm}
\section{Optimal Planning and Operation of a Generalized Multi-Frequency System}\label{sec:OPF_Planning_Operation_Stages}
\begin{figure}[t!]
		\centering
		\includegraphics[width = 1\columnwidth] {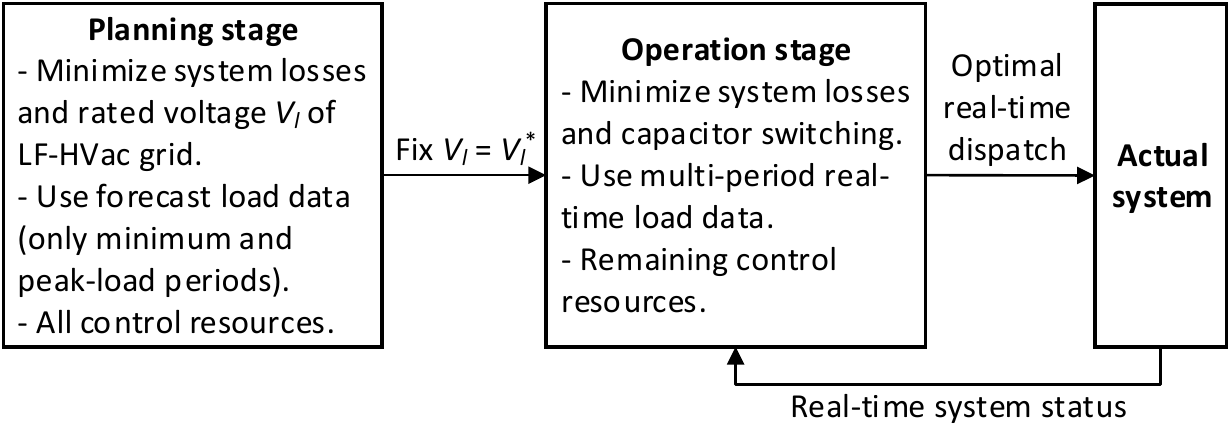}	
		\vspace{-0.7cm}
		\caption{The two optimization problems associated with the planning and operation phases.}		
		\label{fig:Planning_and_Operation_Flowchart}
	\end{figure}

	This section presents two OPF problems for planning and operation stages in a generalized multi-frequency system. The relation between the two problems are shown in Fig. \ref{fig:Planning_and_Operation_Flowchart}.

\vspace{-0.2cm}	
\subsection{Planning Stage: Optimal Voltage of the LF-HVac Grid}\label{subsec:OPF_Planning_Stage}
	In the planning stage, an optimization problem is formulated to determine an optimal transmission voltage rating $V_l$ of LF-HVac grid $l$, given a desired power transfer based on demand forecasting.  Additionally, preliminary operating frequency $F_l$ and dispatch from generators, shunt capacitors, and converters are determined. System losses are minimized by minimizing the total generation in HVac grid $s$ and LF-HVac grid $l$.
	
	A {\it variable vector} $\boldsymbol{X}$ is defined as a combination of state variables $\boldsymbol{x}$ and control variables $\boldsymbol{u}$ as follows:
	
	\vspace{-0.7cm}
	\begin{align}
		\label{eqn:Variables_planning}
		\nonumber
		\boldsymbol{X} = [\boldsymbol{x} | \boldsymbol{u}] = [\boldsymbol{e}_{s}, \boldsymbol{f}_{s}, \boldsymbol{e}_{l}, &\boldsymbol{f}_{l} \hspace{0.1cm} | \hspace{0.1cm} V_l, F_l, {\boldsymbol{P}_{s}^{gen}}, {\boldsymbol{Q}_{s}^{gen}}, {\boldsymbol{Q}_{s}^{sh}},\\
		{\boldsymbol{P}_{l}^{gen}}, {\boldsymbol{Q}_{l}^{gen}}, \boldsymbol{Q}_l^{sh}, & \hspace{0.1cm} {\boldsymbol{P}_{s}^{conv}}, {\boldsymbol{Q}_{s}^{conv}}, {\boldsymbol{P}_{l}^{conv}}, {\boldsymbol{Q}_{l}^{conv}}].
	\end{align}
	A {\it weighted-sum multi-objective function} comprising the total generation and optimum voltage rating of grid $l$ is defined as follows:
	\begin{align}
	\label{eqn:Objectives_Planning}
	f(\boldsymbol{X}) = \alpha_1 \big[\sum_{k \in {\mathcal{G}_s}} \! P_{s,k}^{gen}+\sum_{k \in {\mathcal{G}_l}} \! P_{l,k}^{gen} \big] +  \alpha_2 V_l.
	\end{align}
	{\it Equality and inequality constraints} for grids $s$ and $l$ as well as converters are defined as follows:  
	
	In HVac grid $\boldsymbol{s}$, the equality constraints $\boldsymbol{g}_s$ representing the power balance at every bus $k$ and the voltage magnitude requirement at voltage controlled buses, are given by:
	\begin{align}
		\label{eqn:HVac_equality_constraint}	
		g_{s,k}^{P}(\boldsymbol{X}) & \!=\! P_{s,k} \!-\! (P_{s, k}^{gen} \!-\! P_{s, k}^{load} \!-\! P_{s,k}^{conv}) = 0, \hspace{.1cm} \forall k \in \mathcal{N}_{s},\\
		\nonumber
		g_{s,k}^{Q}(\boldsymbol{X}) &= Q_{s,k} - [Q_{s, k}^{gen} - Q_{s, k}^{load} - Q_{s,k}^{conv} -\\ & \hspace{1.65cm} (e_{s,k}^2+f_{s,k}^2)Q_{s,k}^{sh}] = 0, \hspace{.1cm} \forall k \in \mathcal{N}_{s},\\
		g_{s,k}^{V}(\boldsymbol{X}) &= e_{s,k}^2 + f_{s,k}^2 - {V_{s,k}^{sch}}^2 = 0, \hspace{.1cm}\forall k \in {\mathcal{G}_s \cup \mathcal{V}_s}.
	\end{align}
	The injected power $P_{s,k}$ and $Q_{s,k}$ to grid $s$ at bus $k$ in (3) and (4) are given by:
	\begin{align}
		\label{eqn:AC_injected_power_eqn}
		P_{s,k} = \boldsymbol{G}_{s,k:} (e_{s,k}\boldsymbol{e}_{s} \!+\!  f_{s,k}\boldsymbol{f}_{s}) \!+\!  \boldsymbol{B}_{s,k:} &(f_{s,k}\boldsymbol{e}_{s} \!-\! e_{s,k}\boldsymbol{f}_{s}),\\
		Q_{s,k} = \boldsymbol{G}_{s,k:} (f_{s,k}\boldsymbol{e}_{s} \!-\! e_{s,k}\boldsymbol{f}_{s}) \!-\! \boldsymbol{B}_{s,k:} &(e_{s,k}\boldsymbol{e}_{s} \!+\! f_{s,k}\boldsymbol{f}_{s}),
	\end{align}
	where $\boldsymbol{G}_{s,k:}$ and $\boldsymbol{B}_{s,k:}$ are the the $k^{th}$ row of the conductance and susceptance matrices $\boldsymbol{G_{s}}$ and $\boldsymbol{B_{s}}$. The inequality constraints in HVac grid $s$ represent the lower and upper limits of generators, discrete capacitor dispatch, and load voltages, which are given as follows:
	\begin{align}
		\label{eqn:HVac_inequality_constraint}
		{\ubar{P}_{s,k}^{gen}} &\le {P}_{s,k}^{gen} \le {\bar{P}_{s,k}^{gen}}, \hspace{.1cm}\forall k \in {\mathcal{G}_s},\\
		{\ubar{Q}_{s,k}^{gen}} &\le {Q}_{s,k}^{gen} \le {\bar{Q}_{s,k}^{gen}}, \hspace{.1cm}\forall k \in {\mathcal{G}_s},\\
		Q_{s,k}^{sh} &\in \mathcal{Q}_{s,k}^{sh}, \forall k \in \mathcal{N}_s^{sh},\\
		{\ubar{V}}^{2} \le h_{s,k}^{V}(\boldsymbol{x}) &= e_{s,k}^2 + f_{s,k}^2 \le {\bar{V}}^{2}, \hspace{.1cm}\forall k \in {\mathcal{L}_s}.
	\end{align}
	
	In LF-HVac grid $\boldsymbol{l}$, similar constraints as in (3) - (11) hold. Voltage constraint (11) is imposed to all buses in grid $s$ since it is assumed that they do not to serve any loads. In addition, the rated voltage and operating frequency are constrained as follows:
	\begin{align}
		\label{eqn:frequency_constraint}
		\ubar{F}_l \leq F_l \leq \bar{F}_l,\\
		\ubar{V}_l \leq V_l \leq \bar{V}_l.
	\end{align}		
	For LF-HVac grid $l$, line parameters vary with the rated voltage $V_l$ and operating frequency $F_L$. Therefore, it is important to note that the conductance and susceptance matrices $\boldsymbol{G_{l}}$ and $\boldsymbol{B_{l}}$ in (\ref{eqn:AC_injected_power_eqn}) and (7) are functions of $V_l$ and $F_l$.
	In this planning phase, line current and power limits in both HVac grid $s$ and LF-HVac grid $l$ are not taken into account. Such a treatment allows a tractable solution to determine the optimal rated voltage $V_l$.

	In the BTB converter systems that connects grids $s$ and $l$, as shown in Fig. \ref{fig:Back-to-Back_Converter_Control},  the equality constraints representing the real power balance between $\textup{VSC}_\textup{1}$ and $\textup{VSC}_\textup{2}$, are defined as follows:
	\begin{align}
		\label{eqn:mismatch_conv}
		g_{c,k}^{P}(\boldsymbol{X}) \!=\! P_{l,k}^{conv} \!+\! P_{J2,k} \!+\! P_{sw2,k} \!+\! P_{J1,k} \!+\! P_{sw1,k} \!-\! P_{s,k}^{conv} \!=\! 0,
	\end{align}
	where the Joule loss ($P_{J}$) and switching loss ($P_{sw}$) at either $\textup{VSC}_\textup{1}$ or $\textup{VSC}_\textup{2}$ side are quadratic functions of the corresponding injected/withdrawn power $P^{conv}$ and $Q^{conv}$ as well as the voltage magnitude $V_{s}$ or $V_{l}$ at the ac terminal in grid $s$ or grid $l$, respectively. The explicit form of (\ref{eqn:mismatch_conv}) in terms of these variables in vector $\boldsymbol{X}$ is given as follows \cite{Quan_3}:
	\begin{align}
		\label{eqn:mismatch_conv_final}
		\nonumber
		&g_{c,k}^{P}(\boldsymbol{X})\!=\! P_{l,k}^{conv} \!+\! 2a_0 \!-\! P_{s,i}^{conv}\\
		\nonumber
		&+{{{P_{l,k}^{conv}}^2+{Q_{l,k}^{conv}}^2} \over e_{l,k}^2\!+\!f_{l,k}^2} (R_{2,k} \!+\! a_2) \!+\! \sqrt{{{{P_{l,k}^{conv}}^2 \!+\! {Q_{l,k}^{conv}}^2} \over e_{l,k}^2\!+\!f_{l,k}^2}} a_1\\
		&+ {{{P_{s,k}^{conv}}^2+{Q_{s,k}^{conv}}^2} \over e_{s,k}^2\!+\!f_{s,k}^2} (R_{1,i} \!+\! a_2) \!+\! \sqrt{{{{P_{s,k}^{conv}}^2 \!+\! {Q_{s,k}^{conv}}^2} \over e_{s,k}^2\!+\!f_{s,k}^2}} a_1  \!=\! 0,
	\end{align}
	where $a_0$, $a_1$, and $a_2$ are given coefficients of the loss quadratic function, $R_{1,k}=R_{T1,k}+R_{c1,k}$, and $R_{2,k}=R_{T2,k}+R_{c2,k}$ with $R_{T1,k}$, $R_{T2,k}$, $R_{c1,k}$, and $R_{c2,k}$ being the winding resistances of the transformers and phase reactors.
	
	The inequality constraints that form the feasible operating region for $\textup{VSC}_\textup{1}$ and $\textup{VSC}_\textup{2}$ include:\\
	1) the limit of converter current $I_{c1}$ and $I_{c2}$ to avoid overheating for switching devices,\\
	2) the limit of ac-side converter voltage $V_{c1}$ and $V_{c2}$ by the dc-link voltage $V_{dc}$ to avoid over-modulation, and\\
	3) the limit of reactive power $Q^{conv}$ absorbed by $\textup{VSC}_\textup{1}$ and $\textup{VSC}_\textup{2}$ from HVac grid $s$ and LF-HVac grid $l$, respectively.\\ These constraints are normally converted into equivalent constraints of voltage magnitude and converter power at the ac terminal in order to be easily embedded in optimal power flow algorithms.  Without loss of generality, the following constraints are written using the notation of $\textup{VSC}_\textup{1}$ as follows \cite{Jef_1,Feng_1}:
	\begin{align}
		\label{eqn:converter_Ic_constraint}
		\nonumber
		&1) \hspace{1cm} 
		I_{c1,k} \leq I_{c,k}^{max} \iff   \\
		&{(P_{s,k}^{conv})}^2  \! + \! {(Q_{s,k}^{conv})}^2 \!-\! {(I_{c,k}^{max})}^2 (e_{s,k}^2\!+\!f_{s,k}^2) \!\leq\! 0, \forall k \!\in\! \mathcal{C}_{s}\\
		\label{eqn:converter_Vc_constraint}
		\nonumber
		&2) \hspace{1cm}  
		V_{c1,i} \leq k_m V_{dc,i} \iff \\
		\nonumber
		&[P_{s,k}^{conv} \!-\! (e_{s,k}^2\!+\!f_{s,k}^2)g_{1,k}]^2 + [Q_{s,k}^{conv} \!+\! (e_{s,k}^2\!+\!f_{s,k}^2)b_{1,k}]^2\\
		&\hspace{2cm}  -\bigg(\dfrac{k_mV_{dc}}{Z_{1,k}}\bigg)^2 (e_{s,k}^2\!+\!f_{s,k}^2) \!\leq\! 0, \forall k \!\in\! \mathcal{C}_{s}\\
		\label{eqn:converter_Qconv_constraint}		
		&3) \hspace{1cm}  
		Q_{s,k}^{conv} \leq  k_Q S_{rated, k}^{conv}, \forall k \!\in\! \mathcal{C}_{s},
	\end{align}
	where $k_m$ and $k_Q$ are given coefficients, and where ${\bar{Z}_{1,k}}$ = 1/$(g_{1,k} + jb_{1,k})$ is the combined impedances of the transformer and filters at each VSC side of the BTB converter station. It is important to note that the feasible operating region of a VSC formed by (\ref{eqn:converter_Ic_constraint})-(\ref{eqn:converter_Qconv_constraint}) varies with the ac terminal voltage.

	The properties of the formulated OPF problem during the planning stage and the solution approach are described in Section \ref{sec:Solution_Approach}.

	\vspace{-0.2cm}
\subsection{Operation Stage: Real-Time Operating Frequency of the LF-HVac Grid and OPF in the Multi-Frequency System}	\label{subsec:OPF_Operation_Stage}
	The optimal LF-HVac transmission voltage rating $V^\star_l$ determined in the planning phase above become a given input to the OPF problem during the multi-period operation with real-time power transfer and load levels. LF-HVac grid $l$ optimum operating frequency and generator/converter/shunt capacitor dispatch in both grids $s$ and $l$ are now determined in the operation phase. The OPF formulation follow planning phase with the following important modifications.  
	
	The {\it variable vector} is defined as in (\ref{eqn:Variables_planning}) but without $V_l$ because its value is now known as $V^\star_l$. 
	
	The {\it objective function} of the optimization problem during the operation stage is updated to include the penalty for capacitor switching operations:
	\begin{align}
	\label{eqn:Objectives_Operation}
	\nonumber
	f(\boldsymbol{X}) &= \alpha_1 \big[\sum_{k \in {\mathcal{G}_s}} \! P_{s,k}^{gen}+\sum_{k \in {\mathcal{G}_l}} \! P_{l,k}^{gen} \big] \\
	+  \alpha_3 & \big[\sum_{k \in \mathcal{N}_{s}^{sh}} \!\!\!{(Q_{s,k}^{sh} \!-\! Q_{s,k}^{sh^{pre} })}^{2} \!+\!\! \sum_{k \in \mathcal{N}_{l}^{sh}} \!\!\!{(Q_{l,k}^{sh} \!-\! Q_{l,k}^{sh^{pre} })}^{2} \big],
	\end{align}
	where $Q_{s,k}^{sh^{pre} }$ and $Q_{l,k}^{sh^{pre} }$ are the dispatch of the capacitor at bus $k$ during the previous time step. 
	
	In addition to (3)-(17), the {\it constraints} for HVac grid $s$ and LF-HVac grid $l$ now also include the lower and upper limits of line power and current, which are also written herein only for grid $s$ as follows:
	\begin{align}
		\label{eqn:AC_I_Line_Limits}
		\nonumber
		h_{s,kj}^{I}(\boldsymbol{X}) &= (g_{s,kj}^2+b_{s,kj}^2) [{(e_{s,k}\!-\!e_{s,j})}^2 + {(f_{s,k}\!-\!f_{s,j})}^2] \\
		&\le \bar{I}_{s,kj}^{2},  \forall (k,j) \in {\mathcal{D}_s},\\
		\label{eqn:AC_P_Line_Limits}
		\nonumber
		-\bar{P}_{s,kj} \le& h_{s,kj}^{P}(\boldsymbol{X}) = g_{s,kj} (e_{s,k}^2 \!+\! f_{s,k}^2 \!-\! e_{s,k}e_{s,j} \!-\! f_{s,k}f_{s,j}) \\
		+ b_{s,kj}&(e_{s,k}f_{s,j}\!-\!f_{s,k}e_{s,j}) \le \bar{P}_{s,kj},  \forall (k,j) \!\in \! {\mathcal{D}_s},
	\end{align}
	where ($k, j$) is the line between buses $k$ and $j$. While (\ref{eqn:AC_I_Line_Limits}) is imposed to satisfy the conductor thermal limit, (\ref{eqn:AC_P_Line_Limits}) represents the limit on power transfer to guarantee system stability.
	
\vspace{-0.2cm}	
\section{Problem Challenges and Solution Approach}\label{sec:Solution_Approach}
	It is widely known that solving the OPF problem in conventional 50/60-Hz HVac grid $s$ is challenging because of the nonconvex mixed-integer nonlinear programing (MINLP) characteristic and considerable computational requirement for real-time applications in large transmission systems. To the best of our knowledge, no open-source/commercial solver can handle a large-scale MINLP problem within a reasonable amount of time for real-time applications in power systems. In addition, a general solver does not take advantages of attracting computational properties of OPF constraints in power systems. In this paper, a Python-based tool previously developed for solving MINLP OPF problems in large-scale unbalanced distribution systems \cite{Quan_2} is leveraged with modifications to account for different objectives and constraints of the planning and operation OPF problems at transmission domain. The algorithm in this tool is based on the predictor-corrector primal-dual interior-point (PCPDIPM) method \cite{Torres_1}, which is known for good performance when solving nonconvex optimization problems. The technique to handle the discrete variables in the developed tool is adopted from \cite{Liu_1}.
	
	Considering the formulated objective and constraints in Section \ref{sec:OPF_Planning_Operation_Stages}, the additional challenges of the proposed OPF problems in the planning and operation stages compared to the conventional OPF problems include:\\
	1) varying transmission line parameters in LF-HVac grid $l$ when rated voltage $V_l$ and frequency $F_l$ are control variables, and \\	
	2) varying constraints (\ref{eqn:mismatch_conv_final})-(\ref{eqn:converter_Qconv_constraint}) in the converter model.
	
	To handle the complexity of varying rated voltage and frequency in the OPF model of the LF-HVac grid $l$, instead of using explicit variables $V_l$ and $F_l$ as in Section \ref{sec:OPF_Planning_Operation_Stages}, the ratios $V_r$ and $F_r$ of the rated voltage and frequency to a voltage base and a frequency base are introduced as follows:
	\begin{align}
		\label{eqn:Voltage_Frequency_Ratios}
		F_r = \dfrac{F_l}{F^{base}}, V_r = \dfrac{V_l^2}{{(V^{base})}^2},
	\end{align}
	where $V_{base}$ and $F_{base}$ are chosen to be 500 kV and 60 Hz in this paper. The lumped parameters ($r_{kj}$, $x_{kj}$, $b_{kj}$) in the equivalent-$\pi$ model of a transmission line between buses $k$ and $j$ at voltage base $V_l$ and frequency $F_l$ thus are determined from the given parameters ($r_{kj}^{base}$, $x_{kj}^{base}$, $b_{kj}^{base}$) at voltage base $V^{base}$ and frequency $F^{base}$ as follows:
	\begin{align}
		\label{eqn:New_Transmission_Line_Parameters}
		r_{kj} = \dfrac{r_{kj}^{base}}{V_r}, \hspace{0.1cm}x_{kj} = \dfrac{x_{kj}^{base}F_r}{V_r}, \hspace{0.1cm}b_{kj} = b_{kj}^{base}F_rV_r, \hspace{0.1cm} \forall (k,j) \!\in \! {\mathcal{D}_l}.
	\end{align}	
	In each iteration of an OPF algorithm, these parameters and thus the conductance and susceptance matrices $\boldsymbol{G_{l}}$ and $\boldsymbol{B_{l}}$ are updated to re-calculate constraints (3), (4), (20), and (21) and the corresponding Jacobian and Hessian matrices in LF-HVac grid $l$. Detailed derivations of unfamiliar entries in the Jacobian and Hessian matrices of real power balance constraint (3) for grid $l$ are shown in the Appendix. Although these terms are not constant, the additional computational burden is not high since the number of buses in LF-HVac grid $l$ is significantly less than that in HVac grid $s$.
	
	Regarding the second challenge, the Jacobian and Hessian matrices of variable converter constraints are also derived but not shown explicitly here due to the space limitation.

\vspace{-0.2cm}
\section{Case Study}\label{sec:CaseStudy}
	This section demonstrates the benefits of the proposed formulation in Section \ref{sec:OPF_Planning_Operation_Stages} and solution approach in Section \ref{sec:Solution_Approach} for the planning and operation phases in a multi-frequency HVac system during a 24-hour period. 

\vspace{-0.3cm}
\subsection{System Description}
\begin{figure}[t]
	\centering
	\includegraphics[width = 1\columnwidth] {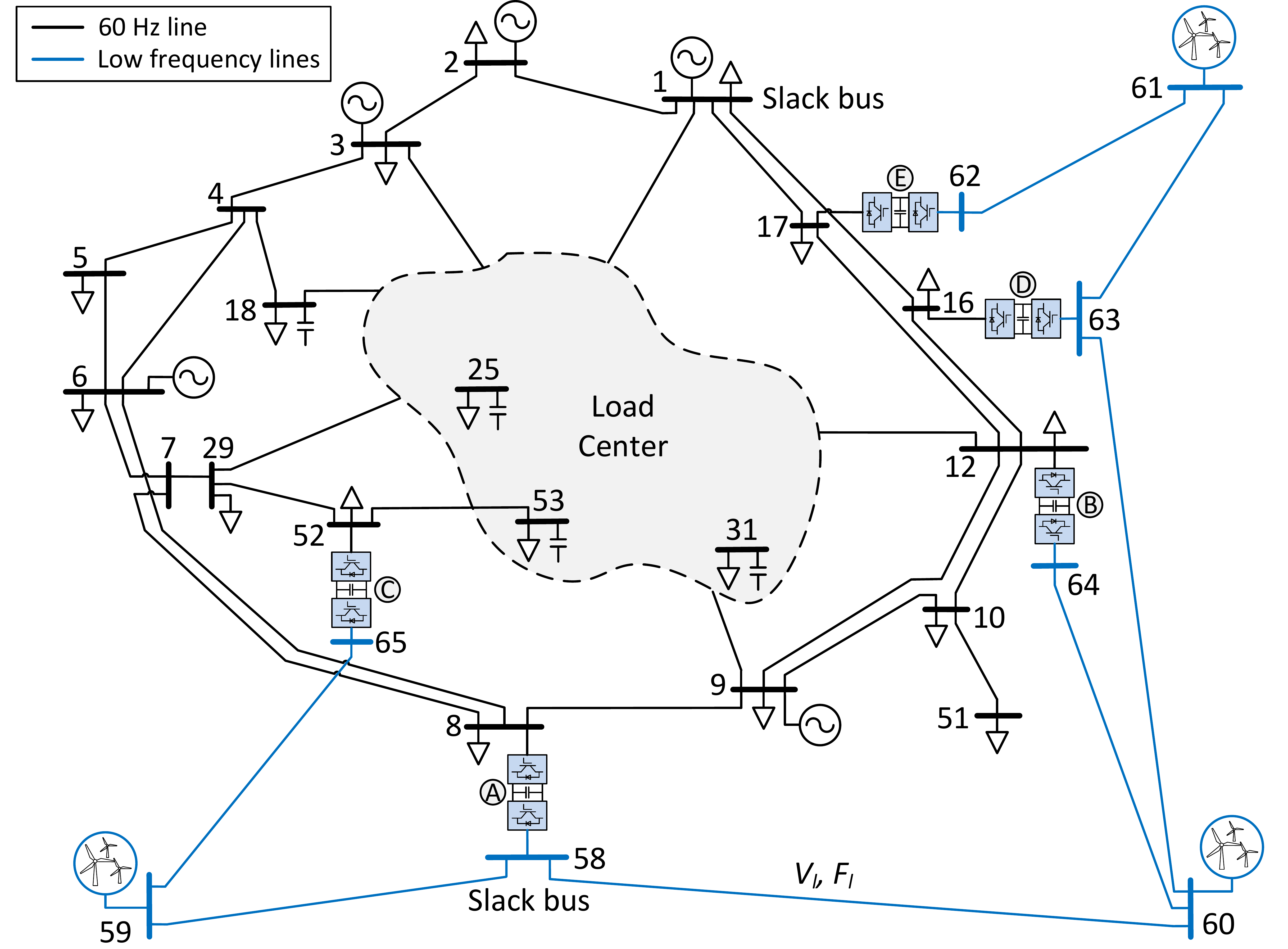}	
	\vspace{-0.5cm}
	\caption{The multi-frequency power system used to validate the proposed optimal power flow formulation and solution. It consists of 60-Hz HVac and LF-HVac lines interconnected by BTB converters. Adapted from \cite{TestTransmissionSystems}.}
	\label{fig:57bus_System}
\end{figure}

	The studied multi-frequency HVac transmission system shown in Fig.~\ref{fig:57bus_System} is modified from the IEEE 57-bus test system \cite{TestTransmissionSystems}. It consists of the original 138 kV 60-Hz HVac grid and a LF-HVac grid. The voltage and frequency of the later are determined as the solution of the proposed planning and operation OPF problems in Section \ref{sec:OPF_Planning_Operation_Stages}. 
	
	The HVac grid consists of 57 buses, including one slack bus, 4 PV buses, and 37 load buses with a peak demand of 1464.3 MW. The 24-hour normalized profile of the actual loads at all buses are assumed to be similar. This grid has 78 transmission lines. Two generators at Bus 8 and 12 in the original grid are replaced by BTB converters A and B. The other BTB converters C, D, and E are connected to Bus 52, 16, and 17, respectively. All of these 5 BTB converters are scheduled to transfer active power from the LF-HVac grid and support reactive power to the HVac grid. Four capacitor banks are located at Bus 18, 25, 31, and 53 as additional reactive power sources with initial dispatches at their maximum rating of 10.01, 9, 10, and 11.88 Mvar, respectively.
	
	The LF-HVac grid consists of 8 buses and 7 300-km transmission lines. There are 5 BTB converters, and the converter at Bus 58, which is the slack bus of the LF-HVac grid, regulates the voltage at this bus. The converters at PV Bus 59-61 are omitted to illustrate an application of wind power generation at low ac frequency.

\vspace{-0.2cm}
\subsection{Planning Stage}
	In the planning stage, a suitable rated voltage $V_l^*$ of the LF-HVac grid $l$ is determined at the minimum and peak load periods. Preliminary optimal frequency $F_l^*$ and dispatch of generators, shunt capacitors, and BTB converters are also identified. The interested ranges of the rated voltage and operating frequency of the LF-HVac grid are [69 - 500] kV and [1 - 60] Hz, respectively. The required voltage limits of load buses in the HVac grid are [0.94 - 1.06] pu.
	
	\begin{table}[b!]
		\setlength{\tabcolsep}{6.3pt}	
		\caption{Solution of the Planning Stage}
		\vspace{-0.2cm}
		\renewcommand{\arraystretch}{1.3}
		\label{tab:Planning_STage_Solution}
		\centering
		\begin{tabular}[h]{|c|c|c|c|c|}
			\hline
			($\alpha_1$, $\alpha_2$)  			& 		$V_l^*$ 		& 	$F_l^*$	& LF-HVac Grid & Total Losses	\\ 	
			& 	($\textup{kV}_\textup{LL}$)											& 	 (Hz)	&  Losses (MW) &  (MW)	\\ 							
			\hline
			\hline
			\multicolumn{5}{|c|}{Minimum load} \\
			\cline{1-5}
			\hline								 	 									   
			(1, 0.0) 								& 357.10			&   14.19		& 	3.62	(0.31\%) & 	13.87	(1.19\%)	\\ 	
			(1, 0.1) 								& 309.70			&   23.38		& 	4.13	(0.35\%) & 	14.72	(1.25\%)	\\
			(1, 0.2) 								& 256.23			&   33.42		& 	5.28	(0.45\%) & 	16.44	(1.40\%)	\\   
			(1, 0.3) 								& 226.93			&   42.80		& 	6.14	(0.52\%) & 	17.82	(1.52\%)	\\  										
			(1, 0.4) 								& 208.82			&   41.73		& 	6.80	(0.58\%) & 	18.92	(1.61\%)	\\  			
			\hline  		
			\hline	
			\multicolumn{5}{|c|}{Peak load} \\	
			\cline{1-5}
			\hline								 	 									   
			(1, 0.0) 								& 351.17			&   14.35		& 	5.45	(0.37\%) & 	21.12	(1.44\%)	\\ 	
			(1, 0.1) 								& 340.89			&   19.77		& 	5.63	(0.38\%) & 	21.25	(1.45\%)	\\
			(1, 0.2) 								& 296.26			&   27.16		& 	7.22	(0.49\%) & 	22.93	(1.57\%)	\\  
			(1, 0.3) 								& 268.44			&   33.61		& 	8.64	(0.59\%) & 	24.47	(1.67\%)	\\  								
			(1, 0.4) 								& 248.63			&   39.84		& 	9.70	(0.66\%) & 	25.89	(1.77\%)	\\  			
			\hline  			
		\end{tabular}
	\end{table}	
	
	Table \ref{tab:Planning_STage_Solution} shows the solution of the proposed planning OPF problem with different set of weighting coefficients ($\alpha_1$, $\alpha_2$) in (\ref{eqn:Objectives_Planning}). When $\alpha_2$, which represents the priority of minimizing the rated voltage of the LF-HVac grid, increases, the resulting optimal voltage $V_l^*$ reduces. In addition, the transmission losses in the LF-HVac grid and the total losses of the multi-frequency system increase significantly less than the square of voltage reduction. Such an achievement results from the optimal dispatch from generators, shunt capacitors, and BTB converters as well as the optimal operating frequency $F_l^*$. It is also important to note that when voltage is not penalized in the objective function, i.e. $\alpha_2$ = 0,  the optimal voltage $V_l^*$  is significantly less than the upper limit 500 kV with the chosen system parameters and loading conditions in this study. At the optimal voltage and operating frequency, the corresponding reactive power consumed and supplied from line reactances, capacitances, and BTB converters result in minimum line current and thus lowest MW losses.

	A rated voltage of 345 kV, which is based on the results corresponding to ($\alpha_1$, $\alpha_2$) = (1, 0.1), is chosen for the operational process in the LF-HVac grid $l$. The analysis of system losses, voltage regulation, and the optimal power dispatch and operating frequency of the LF-HVac grid is described in more detailed in the multi-period operation phase.

	\vspace{-0.2cm}	
\subsection{Operation Stage}
\begin{figure}[t!]
		\centering	
		\includegraphics[width = 0.98\columnwidth] {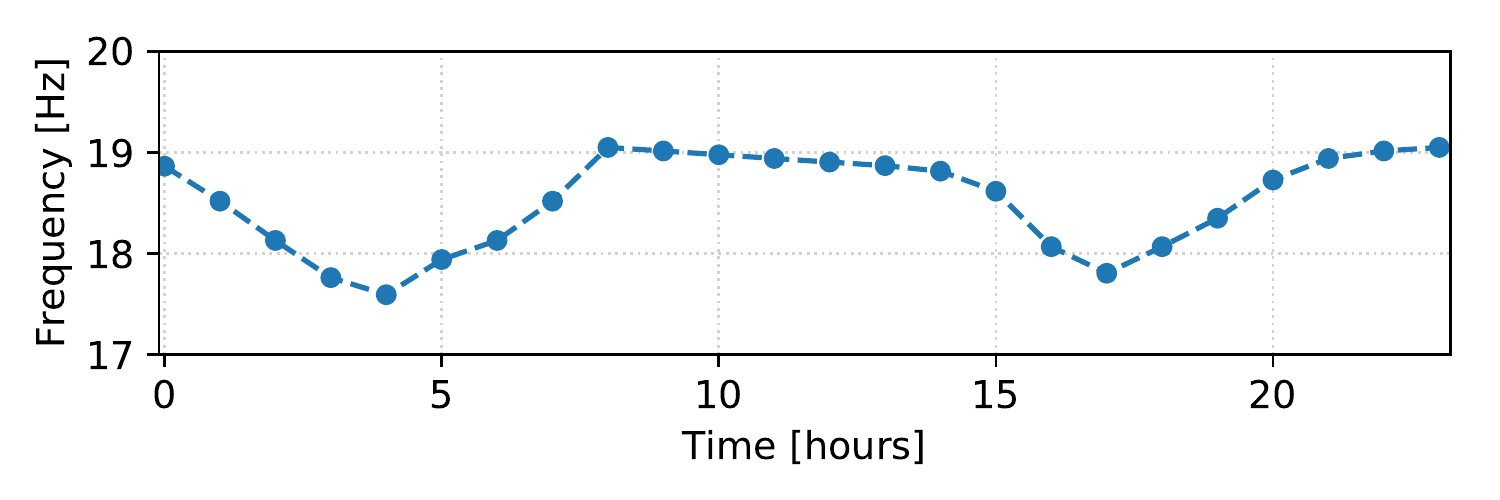}	
		\vspace{-0.3cm}		
		\caption{Optimal operating frequency of the LF-HVac grid.}	
		\label{fig:Optimal_Frequency}
	\end{figure}

	At the chosen voltage level of 345 kV, Fig. \ref{fig:Optimal_Frequency} shows the optimal operating frequency of the LF-HVac grid during the simulated day. The resulting frequency varies within a small range of [17.59 - 19.05] Hz, which is considerably higher than the lower limit of 1 Hz. Table \ref{tab:ConverterDP_3} shows the optimal dispatch at both sides of the BTB converters and the corresponding ac voltages at the points of connection to the HVac and LF-HVac grids during the peak load. The red numbers denote the binding to constraint (18) of converter dispatch. In addition, the dispatch of all capacitors converges exactly at their available discrete values at all time steps. With fast reactive power support from BTB converters, no capacitor switching operations are needed to regulate load voltage with respect to demand variations.
	\begin{table}[t!]     
		\setlength{\tabcolsep}{4pt}	                                                 
		\vspace{-0.1cm}	
		\caption{Optimal Power Dispatch of BTB Converters in Case 3 at 5 pm.}
		\vspace{-0.2cm}
		\renewcommand{\arraystretch}{1.3}
		\label{tab:ConverterDP_3}
		\centering
		\begin{tabular}[h]{|c|ccc|ccc|}
			\hline
			& \multicolumn{3}{|c|}{From HVac to VSC$_1$}  							& \multicolumn{3}{|c|}{From VSC$_2$ to LF-HVac} 										\\ \cline{2-7} 
			Converters  & {$P_s^{conv}$}   				& {$Q_s^{conv}$} 			& {$V_s$}	  	& {$P_l^{conv}$}			& {$Q_l^{conv}$} 			& {$V_l$} 		\\
						& {(MW)}   						& {(Mvar)} 					&{(pu)}  		& {(MW)}					& {(Mvar)} 					&{(pu)}    		\\		
			\hline								 	 									   
			$A$         & -231.11 						& -28.31    				& 0.9829	 	& -231.87					&   56.82					&  1.02			\\ 
			$B$         & -296.28  						& -44.17					& 0.9966		& -297.32					&  -27.43   				&  1.0059 		\\ 
			$C$         & -35.83  						& -5.85    					& 1.0395		& -36.08					&  \textcolor{red}{-20.0}	&  1.0088 		\\ 
			$D$         & \textcolor{red}{-200.21}		& \textcolor{red}{-43.77}	& 1.0237		& -200.84  					&  \textcolor{red}{-20.0}   &  1.0096		\\ 
			$E$         & -147.83  						& -54.34	     			& 1.0347		& -148.30					&  -14.76    				&  1.0079		\\ 	
			\hline  		
		\end{tabular}
	\end{table}

\begin{figure}[t!]
		\centering	
		\includegraphics[width = 0.98\columnwidth] {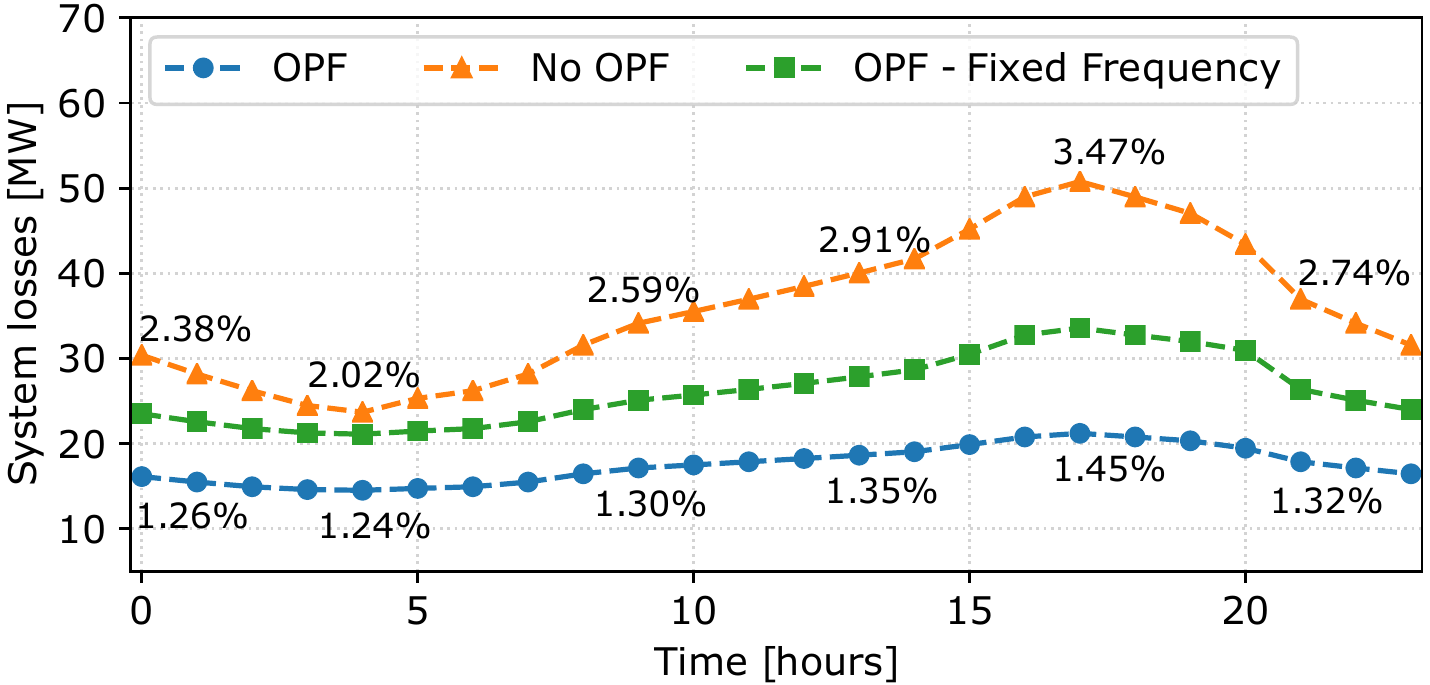}	
		\vspace{-0.2cm}		
		\caption{System losses in the simulated day without and with optimal control.}			
		\label{fig:Losses}
	\end{figure}

	Fig. \ref{fig:Losses} shows the MW losses of the multi-frequency power system and the corresponding percentages compared to the demand in three cases. In Case 1, OPF is disabled, and the system operating points are based on a given power dispatch of generators, shunt capacitors, and BTB converters and a PF solver developed for multi-frequency system \cite{Quan_3}. Case 2 corresponds to an enabled OPF with a fixed operating frequency of 5 Hz in the LF-HVac grid. In Case 3, all dispatch resources, i.e. from generators, shunt capacitors, and BTB converters as well as the operating frequency of the LF-HVac grid are considered as control resources of the OPF. In both Cases 2 and 3, the capacitor switching operations are penalized by choosing ($\alpha_1, \alpha_3$) = (1, 0.2). The system losses follows the load profile, and it is less sensitive to load variations with optimal control. The losses are highest when OPF is disabled in Case 1 and lowest in Case 3 with all optimal control resources. At the peak load, the system losses reduce from 3.47\% in Case 1 to 2.29\% and 1.45\% in Cases 2 and 3, respectively. Although the losses in Case 2 is less than that in Case 1, it is still significantly higher than the losses in Case 3. Similarly, when a fixed dispatch is assigned for all BTB converters and the operating frequency is variable, similar total losses are observed as in Case 2. These results show the importance of including the BTB converter dispatch and operating frequency of the LF-HVac grid as control resources in the proposed operation OPF in addition to conventional generators and shunt capacitors. 
	
\begin{figure}[t!]
		\centering	
		\includegraphics[width = 0.97\columnwidth] {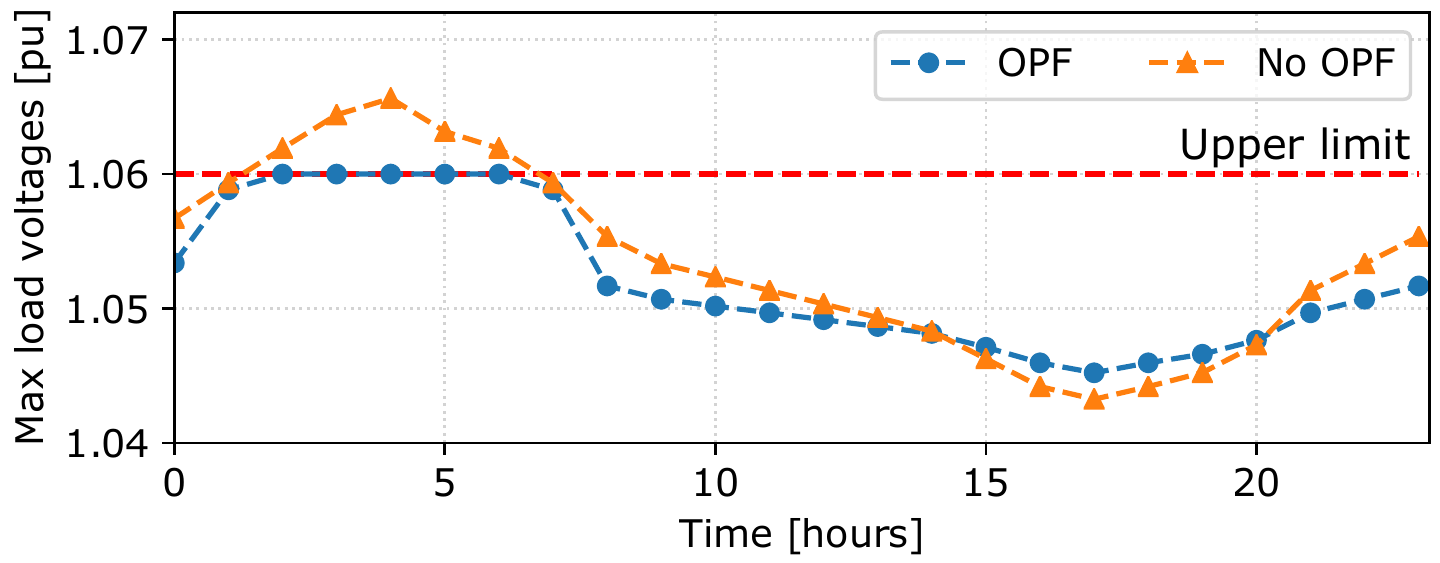}	
		\vspace{-0.2cm}		
		\caption{Highest load voltages in the three cases in the simulated day.}
		\label{fig:Vload_max}
	\end{figure}

	
	Fig. \ref{fig:Vload_max} shows the highest voltages at all load buses in the studied cases during the simulated day. While overvoltage appears when OPF is disabled in Case 1, the optimal dispatch of the generators, shunt capacitors, and BTB converters and the operating frequency of the LF-HVac grid eliminate voltage violation in the system through out the day in Case 3.
	

	\vspace{-0.3cm}
\subsection{Convergence and Scalability}
	The proposed solution approach in Section \ref{sec:Solution_Approach} is implemented in Python. The convergence tolerance is set to $10^{-6}$ pu for solution feasibility, objective function, and complementary gap. A flat start is used in solving the OPF problem the planning stage, while both a flat start and preliminary solution obtained from the planning stage are used in the OPF problem for the operation phase. It is observed that these starting point strategies perform well in terms of solution accuracy, run time, and number of iterations. Both flat and warm starts lead to identical solutions in spite of the nonconvexity of the formulated OPF. The average run times per step when solving the OPF problems in the planning and operation stages are approximately 3.5 and 4.6 seconds, respectively. The longer run time in the operation phase results from the additional line current and power constraints (\ref{eqn:AC_I_Line_Limits}) and (\ref{eqn:AC_P_Line_Limits}).

	\vspace{-0.2cm}
\section{Conclusion}
	This paper proposes an optimal planning and operation for a multi-frequency HVac transmission system. The losses in the LF-HVac grid, which includes transmission and converter losses, depends on the operating voltage, frequency, and converter dispatch. Therefore, multi-objective OPF-based formulations and solution approach for both planning and operation stages are formulated. The results of the planning phase show different optimal rated voltages of the LF-HVac grid corresponding to different weighting factors between total losses and the rated voltage in the objective function. The results of the operation phase show that the optimal frequency of the LF-HVac grid varies within a small range between 17.5 and 19 Hz during the simulated day with a chosen load profiles. More importantly, the operating frequency and converter dispatch show significant impacts on reducing system losses up to 2\% during the peak-load condition and elimination of voltage violations. The solution of the planning and operation stages converge in all time step with acceptable run time and number of iterations, which shows the possibility of applications in larger multi-frequency HVac power systems, including connections to HVdc grids.

	\vspace{-0.2cm}
\section{Appendix}
	This Section shows the generalized derivations to calculate the unfamiliar entries in the Jacobian and Hessian matrices corresponding to the real power balance constraint (3) in the LF-HVac grid, as described in Sections \ref{sec:OPF_Planning_Operation_Stages} and \ref{sec:Solution_Approach}. Because of space limitation, only first and second derivatives with respect to frequency ratio $F_r$ are shown. The subscript $l$ and superscript $base$ that signifies grid $l$ and line parameters ($r_{kj}^{base}$, $x_{kj}^{base}$, $b_{kj}^{base}$) are dropped for simplicity. \vspace{-0.3cm}

	\small
	\begin{align}
		\nonumber
		&\frac{\partial g_{k}^P}{\partial F_r} = \frac{\partial}{\partial F_r} \bigg(\boldsymbol{G}_{k:} (e_{k}\boldsymbol{\bar{e}} + f_{k}\boldsymbol{\bar{f}}) + \boldsymbol{B}_{k:} (f_{k}\boldsymbol{\bar{e}} - e_{k}\boldsymbol{\bar{f}})\bigg)\\
		\nonumber
		&= (e_{k}^{2} \!+\! f_{k}^{2}) \frac{\partial G_{kk}}{\partial F_r}  \!+\!
		\sum_{j \neq k} \! \bigg[(e_{k}e_{j} \!+\! f_{k}f_{k})\frac{\partial G_{kj}}{\partial F_r} \!+\! (f_{k}e_{j} \!-\! e_{k}f_{k})\frac{\partial B_{kj}}{\partial F_r} \boldsymbol{} \bigg]\\
		\nonumber		
		&\frac{\partial^2 g_{k}^P}{\partial^{2} F_r}  = \frac{\partial}{\partial^{2} F_r}  \bigg(\boldsymbol{G}_{k:} (e_{k}\boldsymbol{\bar{e}} + f_{k}\boldsymbol{\bar{f}}) + \boldsymbol{B}_{k:} (f_{k}\boldsymbol{\bar{e}} - e_{k}\boldsymbol{\bar{f}})\bigg)\\
		\nonumber
		&=\! (e_{k}^{2} \!\!+\!\! f_{k}^{2}) \frac{\partial^{2} G_{kk}}{\partial^{2} F_r} 
		\!+\! \sum_{j \neq k} \! \bigg[ (e_{k}e_{j}\!\!+\!\!f_{k}f_{k})\frac{\partial^{2} G_{kj}}{\partial^{2} F_r} \!+\! (f_{k}e_{j}\!-\!e_{k}f_{k})\frac{\partial^{2} B_{kj}}{\partial^{2} F_r}\bigg]\\
		\nonumber		
		&\frac{\partial^2 g_{k}^P}{\partial F_r \partial e_{j}} = \frac{\partial^2 g_{k}^P}{\partial e_{j} \partial F_r} = \boldsymbol{} 
		\begin{cases}
		2e_{k}\frac{\partial G_{kk}}{\partial F_r} + e_{j}\frac{\partial G_{kj}}{\partial F_r} - f_{k}\frac{\partial B_{kj}}{\partial F_r}, \hspace{0.2cm} \text{if $j=k$}\\[3pt]
		e_{k}\frac{\partial G_{kj}}{\partial F_r} + f_{k}\frac{\partial B_{kj}}{\partial F_r}, \hspace{0.2cm} \text{otherwise}.
		\end{cases}\\
		\nonumber		
		&\frac{\partial^2 g_{k}^P}{\partial F_r \partial f_{k}} = \frac{\partial^2 g_{k}^P}{\partial f_{k} \partial F_r} = 
		\nonumber
		\begin{cases}
		2f_{k}\frac{\partial G_{kk}}{\partial F_r} + f_{k}\frac{\partial G_{kj}}{\partial F_r} + e_{j}\frac{\partial B_{kj}}{\partial F_r}, \hspace{0.2cm} \text{if $j=k$}\\[3pt]
		f_{k}\frac{\partial G_{kj}}{\partial F_r} - e_{k}\frac{\partial B_{kj}}{\partial F_r}, \hspace{0.2cm} \text{otherwise}.
		\end{cases}\\
		\nonumber
		&\frac{\partial^2 g_{k}^P}{\partial F_r \partial V_r} = \frac{\partial^2 g_{k}^P}{\partial V_r \partial F_r} = (e_{k}^{2} \!+\! f_{k}^{2}) \frac{\partial^{2} G_{kk}}{\partial F_r \partial V_r}\\
		\nonumber
		&\hspace{1.5cm}\!+\! \sum_{j \neq k} \! \bigg[ (e_{k}e_{j}\!+\!f_{k}f_{k})\frac{\partial^{2} G_{kj}}{\partial F_r \partial V_r} \!+\! (f_{k}e_{j}\!-\!e_{k}f_{k})\frac{\partial^{2} B_{kj}}{\partial F_r \partial V_r}\bigg]
	\end{align}
	where:
	\begin{align}
		\nonumber
		\frac{\partial G_{kj}}{\partial F_r} &\!=\!\frac{2r_{kj}x_{kj}^{2}F_rV_r}{(r_{kj}^{2}+x_{kj}^{2}F_r^{2})^{2}}, \hspace{0.2cm} \frac{\partial B_{kj}}{\partial r} = \frac{x_{kj}(r_{kj}^{2}-x_{kj}^{2}F_r^{2})V_r}{(r_{kj}^{2}+x_{kj}^{2}F_r^{2})^{2}}\\
		\nonumber
		\frac{\partial G_{kk}}{\partial F_r} &\!=\! - \sum_{j \neq k}\frac{\partial G_{kj}}{\partial F_r}, \hspace{0.2cm}
		\frac{\partial B_{kk}}{\partial F_r} = \sum_{j \neq k} \bigg[ -\frac{\partial B_{kj}}{\partial F_r} + \dfrac{b_{kj}V_r}{2} \bigg],\\
		\nonumber
		\frac{\partial^2 G_{kj}}{\partial^2 F_r} &\!=\! \frac{2r_{kj}^{3}x_{kj}^{2}\!-\!6r_{kj}x_{kj}^{4}F_r^{2}}{(r_{kj}^{2} \!+\!x_{kj}^{2}F_r^{2})^{3}}, 
		\frac{\partial^2 B_{kj}}{\partial^2 F_r} \!=\! \frac{2(x_{kj}^{5}F_r^{2} \!-\! 3r_{kj}^{3}x_{kj}^{2})F_rV_r }{(r_{kj}^{2}+x_{kj}^{2}F_r^{2})^{3}},		\\
		\nonumber
		\frac{\partial^2 G_{kk}}{\partial^2 F_r} &\!=\! -\sum_{j \neq k}\frac{\partial^2 G_{kj}}{\partial^2 F_r}, \hspace{0.2cm} \frac{\partial^2 B_{kk}}{\partial^2 F_r} = -\sum_{j \neq k}\frac{\partial^2 B_{kj}}{\partial^2 F_r}, \hspace{0.2cm} \forall (k,j) \!\in \! {\mathcal{D}_l}
	\end{align}	

\bibliography{MFAC_Optimal_Frequency}
\bibliographystyle{IEEEtran}

\end{document}